\documentclass[aps,preprint]{revtex4-1}
\usepackage{graphicx}
\usepackage{array}
\newcolumntype{+}{>{\global\let\currentrowstyle\relax}}
\newcolumntype{^}{>{\currentrowstyle}}
\newcommand{\rowstyle}[1]{\gdef\currentrowstyle{#1}%
#1\ignorespaces}
\begin{document}

\title{\Large Strong Gravity  Effects of Rotating Black Holes: Quasiperiodic Oscillations}
\author{\large Alikram N. Aliev}
\address{Yeni Y\"{u}zy{\i}l University, Faculty of Engineering and Architecture, Cevizliba\v{g}-Topkap{\i}, 34010  Istanbul, Turkey}
\author{\large G\"{o}ksel Daylan  Esmer}
\address{Istanbul University, Department of Physics,
Vezneciler, 34134 Istanbul, Turkey}
\author{\large Pamir Talazan}
\address{T\"{U}B\.{I}TAK-B\.{I}LGEM, 41470  Gebze, Kocaeli,  Turkey}

\date{\today}

\begin{abstract}

We explore strong gravity effects of the geodesic motion in the spacetime of  rotating black holes in general relativity and braneworld gravity. We  focus on the description of the motion in terms of three fundamental frequencies: The {\it orbital}  frequency, the {\it radial} and  {\it vertical} epicyclic frequencies. For a Kerr black hole, we perform  a detailed numerical analysis of these frequencies  at the innermost stable circular orbits and beyond them as well as at the characteristic stable orbits, at which the radial epicyclic frequency attains its highest value. We find that the values of  the epicyclic frequencies for a class of stable orbits  exhibit good  qualitative agreement with the observed frequencies of the twin peaks quasiperiodic oscillations (QPOs) in some black hole binaries. We also find that at the characteristic stable circular orbits, where  the radial (or the vertical) epicyclic frequency has maxima, the vertical and radial epicyclic frequencies exhibit an approximate  $ 2 : 1 $ ratio even  in the case of near-extreme  rotation of the black hole. Next, we perform  a similar analysis of the fundamental frequencies for a rotating braneworld black  hole and argue that  the existence of such a black hole with a negative tidal charge, whose angular momentum exceeds the Kerr bound in general relativity, does not confront with the observations of high frequency QPOs.

\end{abstract}

\pacs{04.20.Jb, 04.70.Bw, 04.50.+h}

\maketitle

\section{Introduction}

Black holes are one of the most exciting objects of study in modern theoretical physics and astrophysics. They occupy a central place in  all theories of gravity  formulated in various spacetime dimensions.

\subsection{Black holes in general relativity and braneworld gravity}

In four dimensions, general relativity (GR) admits a family of stationary black hole solutions  which turns out to be crucial for understanding its nature and the occurrence of spacetime singularities \cite{penrose}. These solutions possess a number of striking properties such as uniqueness and stability, hidden symmetries and integrability of geodesics etc. \cite{Israel, hawking1, carter1} (see also \cite{chandra, fn}). Altogether, these properties pave the way for astrophysical implications of GR in the regime of strong gravity, stimulating  the search for black holes in the physical universe.

An astrophysical black hole is thought to be described by an exact solution of GR, discovered by  Kerr \cite{kerr}.  This  solution is  uniquely characterized by two parameters: The mass and the angular momentum. In \cite{carter1}, it was first shown that the Hamilton-Jacobi equation for geodesics in the Kerr metric admits a complete separation of variables that is, the geodesic motion is completely  integrable in this spacetime. In subsequent developments, various aspects of the geodesic motion in the Kerr metric have been extensively explored by many authors (for instance, see  \cite{chandra} and references therein). Among these results, the calculations of the observable orbits of test particles are of particular interest. The authors of  \cite{bardeen}  were the first  to give a complete  description of these orbits for the circular motion in the equatorial plane of the Kerr black hole. In particular, they found that for an extreme  rotating  Kerr black hole, the binding energy  of the innermost stable circular orbit (ISCO) (the  maximum amount of energy released by a  particle approaching this orbit)  can attain nearly 42\% of the particle rest-mass energy, whereas for a Schwarzschild black hole it is about 6\%. The high efficiency of this process led to the idea of invoking  an accretion disk around a black hole to explain enormous energy output from both  X-ray binaries and active galactic nuclei (for details, see \cite{shapiro}).

In higher dimensions, there exist a  number of important black hole solutions with unexpected physical characteristics, such as dynamical instability, different horizon topologies and  different rotation dynamics. Among them, the most interesting  black hole solution  is given by the Myers-Perry  metric \cite{mp}, which is supposed to be a generalization of the Kerr solution to all number of spacetime dimensions.  However,  this solution is not unique, unlike the Kerr solution. In five-dimensional vacuum gravity there exists a rotating {\it black ring} solution \cite{er1}, which may have the same mass and angular momentum as the Myers-Perry solution.

On the other hand, the higher-dimensional black holes may have observable consequences as well. We recall that some phenomenological aspects of {\it string/M} theory  lead to  braneworld scenarios, which imply that our physical universe is geometrically described  by a (3+1)-dimensional hypersurface (a ``3-brane")  embedded in higher dimensions \cite{add,rs}.  A complete description of black holes in such scenarios is  a thorny  issue and many aspects of it still remain open. Remarkably, there are several simplifying approaches which provide an intriguing description. For instance, if the size of the  black hole is much smaller than the scale of the  extra dimensions, the black hole would behave as a higher-dimensional object. To good enough approximation, one can argue that these black holes  are described by  Myers-Perry type solutions of the Einstein field equations in higher dimensions.  Furthermore, the fact that in the braneworld scenarios the scale of quantum gravity  becomes as lower as TeV-energy scales indicates  that the small-size black holes would have been detected at high energy experiments \cite{gt, galtsov}.  Of course, there always exists the question  of gravitational stability  of the Myers-Perry type black holes on the brane, whose final status is still not clear. Some interesting aspects of this issue were investigated in  \cite{berti}.

As for the large black holes, for which the size of the horizon is much larger than the length scale of  the extra dimensions, they would also carry the imprints of the extra dimensions though ``effectively" they must look like four-dimensional objects. In one approach employed in \cite{chr}, it was suggested that  such a  black hole in the Randall and Sundrum (RS) braneworld scenario could be described by the ordinary Schwarzschild  solution on the brane,  which  in five dimensions  would look  like  a {\it black string}. The main drawback here is that  the black string solution  suffers from curvature singularities, propagating along the extra dimension. This solution also suffers from   classical Gregory-Laflamme instabilities \cite{gl}.  In another  approach,  such black holes  were described by  postulating the metric form and  solving the  effective gravitational equations on the RS brane \cite{dmpr, ae2}.  This idea is also reinforced  by the fact that in the low energy limit, the  RS braneworld  scenario supports the properties of four-dimensional GR  on the brane \cite{sms1, ae1}. It is interesting that in this approach one can analytically describe, to some extent,  the braneworld  black holes  by a Reissner-Nordstrom type solution in the static case \cite{dmpr} and by a Kerr-Newman type solution in the rotating case \cite{ae2}. The tidal charge appearing in both solutions and being the imprint of the extra fifth dimension supersedes the square of  the electric charge in the usual Reissner-Nordstrom and Kerr-Newman  solutions. Furthermore, the tidal charge in these solutions may have  both {\it positive} and {\it negative} values.  On a par with the Kerr black hole,  the rotating braneworld black hole can be  considered as the useful and the simplest model to explore the physical effects of the tidal charge, though the gravitational stability of the latter  remains  an open problem. Some astrophysical implications of  the braneworld black holes have recently  been  discussed in \cite{stuch1, stuch2, ap, bambi}.

\subsection{High-frequency quasiperiodic oscillations}

Nowadays, astronomical observations provide compelling  evidence for the existence of black holes in the universe.  To focus only on black holes in X-ray binaries, today 23 black holes are known  and  32  X-ray binaries are thought to contain black hole candidates \cite{fer}. Observations of the X-ray binaries with both black holes and black hole candidates have  revealed  finite-width  peaks in the X-ray spectra, which  can be thought of as  signatures of  high-frequency ($>$ 40 Hz) {\it quasiperiodic oscillations} (QPOs) of the black hole accretion disk. Furthermore, in a number of sources, two  peaks of QPOs  have been detected. The associated frequencies of these peaks are (300, 450 Hz) for  X-ray binary GRO J1655-40; (113, 168 Hz) for X-ray binary GRS 1915 + 105;  (184, 276 Hz) for X-ray binary XTE J1550-564 and  (165, 241 Hz) for X-ray binary H 1743-322, harboring a black hole candidate.  The twin frequencies  are in a 3:2 ratio. The source GRS 1915 + 105  also exhibits a pair  (164, 328 Hz) which is in  a 2:1 ratio (see \cite{remi2} for further details).

Observations  of such  frequencies  are certainly of fundamental importance as they  may provide crucial probes of strong gravity near the black holes. Moreover, when combined with the measurements of  the angular momenta of black holes \cite{gou,nara}, they  raise once again  the old compelling question: {\it What is the bona fide geometry of the spacetime around  the black holes}?  For instance,  as argued  in \cite{nara},  the lower bound  on the angular momentum  of a black hole in the X-ray binary GRS 1915 + 105 turns out to be very close to the (upper) Kerr bound  in GR. This raises the question: {\it Does the Kerr solution of GR describe the observed black holes with the high value of the angular momentum}? Today it remains unknown how  future observational data will deal with these questions.  However, this perspective in itself  is very challenging  for  further  study of observable    effects of black holes in both  GR and beyond it, including the cases with the violation of the Kerr bound. Some effects of rotating  black holes in string theory, for which the Kerr bound is breached, were recently studied in \cite{gimon}.

It is also interesting that in a number of  theoretical models explaining the origin of high frequency QPOs,  general relativistic epicyclic frequencies of the geodesic motion  have been considered as `` building blocks", though the precise mechanism  still remains unknown \cite{stella, abram1, abram2, remi2}. The successive theory of the epicyclic motion  around the Kerr black holes in general relativity (with an electric charge or  with an external uniform magnetic field) was  developed in \cite{ag1} (see also companion papers \cite{ag2, ag3, ag4}). The authors,  for the first time, obtained  closed analytical expressions for the epicyclic frequencies of radial and vertical oscillations and  put forward an idea of nonlinear resonances, which may occur when these  frequencies  are in a rational relation \cite{ag1}.  Afterwards,  analytic expressions for the frequency of radial and vertical oscillations were also appeared in \cite{kato1, kato2} in the context of the analysis of trapped oscillations of an accretion  disk around the Kerr black hole. The theory of epicyclic motion and the idea of  nonlinear resonances have been  utilized  in \cite{abram1, abram2}, even reproducing in some cases the plots of \cite{ag1} for the positions of  nonlinear resonances, to explain the observed frequency ratios  of high frequency  QPOs in black hole binaries.

\subsection{Motivation and the basic idea}

In light of all these developments, it  becomes of interest to return once more to the exploration of  the observable effects of a strong gravity regime encoded in the geodesic  motion around  rotating black holes.  Our motivation for this arises from the fact that the epicyclic frequencies of the geodesic motion, as mentioned above, lie at the root of  theoretical models for the observed high frequency  QPOs in black hole binaries. Furthermore, the continuing success with  the measurements  of the angular momentum of  black holes  motivates us to explore the effects of the geodesic  motion not only for rotating black holes in GR, where the Kerr bound on the angular momentum holds, but also for rotating black holes in a braneworld, for which the Kerr bound may be breached. The basic idea of the present paper is as follows:

(i)  To  perform  a detailed numerical analysis of the  epicyclic frequencies  at the stable circular orbits of interest in the Kerr field, using the  general analytical expressions found in earlier papers \cite{ag1,ag2}. Thus, ``monitoring" the behavior  of these frequencies at the stable orbits, especially at those where the radial (or the vertical) epicyclic frequency attains its  highest value, we intend  to clarify the question: {\it Whether the simple model of pure general relativistic geodesic  motion can provide observationally viable values for the epicyclic frequencies  and their ratios}?

(ii) To  extend the numerical analysis  of the epicyclic frequencies to the case of  a rotating braneworld black hole as well. The fact that the associated spacetime metric  has the  Kerr-Newman type form with the  tidal charge (instead of the square of the electric charge) and  for the negative tidal charge, the angular momentum of the black hole exceeds the Kerr bound
makes this model very instructive. It is also important to note that the circular motion in this model may occur in stable orbits, unlike the case of the ordinary black holes in higher dimensions, for instance, the case of Myers-Perry black holes \cite{frs}. Exploring the behavior of the epicyclic frequencies at particular orbits of interest and comparing the results with those for the Kerr black hole we want to address the issues: {\it How the presence of the tidal (positive or negative) charge influences the values of  the epicyclic frequencies? Does the potential existence of over-rotating braneworld black holes  confront with modern observations of high frequency  QPOs in black hole binaries}?

The organization of the paper is as follows: In Sec.II we begin by a brief review of the salient  features of the  geodesic motion  in the Kerr metric  and  give  analytic expressions for the epicyclic frequencies of radial and vertical oscillations. Next, we present the results of a detailed numerical analysis of these frequencies at the stable circular orbits of interest.
In Sec.III we  briefly discuss the salient  properties of a rotating black hole, localized  on  a 3-brane in the Randall-Sundrum  scenario as well as those of the  geodesic motion in the field of this black hole. Here we present analytic expressions for the epicyclic frequencies of radial and vertical oscillations and perform their full numerical analysis, focusing on the stable orbits, where  the radial epicyclic frequency reaches  its maximum value.  We also  discuss the observational signature of the negative tidal charge carried by the braneworld black hole and argue that the idea that such a black hole can  potentially exist in the universe does not confront  with the observations of high frequency QPOs in black holes binaries.  In Sec.IV  we conclude with a  brief discussion of our new results.

\begin{center}
\section{Geodesic Motion and Epicyclic Frequencies in the spacetime  of  Kerr Black Holes}
\end{center}

We begin with the Kerr spacetime which in the Boyer-Lindquist coordinates is given by the metric
\begin{equation}
ds^2  =  -{{\Delta}\over {\Sigma}} \left(dt - a \sin^2\theta\,
 d\phi\,\right)^2 + \Sigma \left(\frac{dr^2}{\Delta} + d\theta^{2}\right) + \frac{\sin^2\theta}{\Sigma}
\left[a dt - (r^2+a^2) \,d\phi \right]^2 \,,
\label{kmetric}
\end{equation}
where
\begin{eqnarray}
\Delta &=& r^2 + a^2 -2 M r\,,~~~~~~
\Sigma= r^2+a^2 \cos^2\theta \,,
\label{kmfunc}
\end{eqnarray}
and $ M $ is  the mass, $ a $ is the rotation parameter or the angular momentum per unit mass, $ a=J/M $.  The most striking  feature  of this spacetime  is that it  contains a rotating black hole.  This  becomes evident by the existence of the event horizon which is governed by the equation $ \Delta =0 $. The largest root of this equation given by
\begin{equation}
r_{+}= M + \sqrt{M^2 - a^2 }\,
\label{horizon11}
\end{equation}
determines the radius of the horizon.  The event horizon, as follows from this expression, exists provided that  $ a \leq M\, $. Thus, {\it a  rotating black hole in general relativity must possess an angular momentum not exceeding its mass}.

Another remarkable feature of the Kerr spacetime is that it admits  the complete separation of variables in the Hamilton-Jacobi equation for geodesics. The global time-translational and rotational symmetries of this spacetime,  along with its hidden symmetries,  make  the Hamilton-Jacobi equation completely integrable \cite{carter1} (see also \cite{chandra}). For our purposes in the following, we will focus only on the  motion  of  test particles  in the equatorial plane $ \theta=\pi/2 $. From the symmetry considerations, it follows that such a motion will occur in circular (cyclic) orbits.  Following works of \cite{ag1, ag2, ag3, ag4}, one can simply describe the circular  motion and its perturbations, i.e. off-equatorial (epicyclic) motion, using the method of successive approximations in the geodesic equation
\begin{equation}
\frac{d^2 x^{\mu}}{ds^2}+ \Gamma^{\mu}_{\alpha \beta} \frac{d
x^{\alpha}}{ds}\frac{d x^{\beta}}{ds}= 0 \,,
\label{eqmot}
\end{equation}
where the  parameter  $ \,s \,$  denotes the proper time along the geodesics and  $ \Gamma^{\mu}_{\alpha \beta} $  are the Christoffel symbols of  the spacetime metric. Clearly, for the circular motion in the equatorial plane  $  r=r_0, \,\,\theta=\pi/2 $,\,  one can introduce the position vector $ z^{\mu}(s) = \{t(s) \,\,,r_0\,\, ,\pi/2\,\,,\Omega_0 t(s) \},\,$where $ \Omega_0 $ is  the orbital frequency. Next, using the Christoffel symbols  for the Kerr metric  it is easy to show that for the circular motion, the $ \mu=0\,,2\,,3 \,$ components of  equation (\ref{eqmot}) become trivial, whereas the remaining component with  $ \mu=1 $ yields
\begin{equation}
\Omega_0 =\frac{\pm\, \Omega_s}{1 \pm a \Omega_s}\,\,.
\label{orbf}
\end{equation}
where $ \Omega_s = M^{1/2}/r^{3/2} $ is  the usual Kepler frequency, the upper sign refers to direct orbits (the motion of the particle and the rotation of the black hole occur in the same direction) and  the lower sign refers to retrograde orbits (the motion of the particle is opposite to the rotation of the black hole). Using equation (\ref{orbf}) in the normalization condition  for the four-velocity, $\,g_{\mu\nu} u^{\mu} u^{\nu}= -1\,$, we obtain the expression for energy of the particle
\begin{equation}
\frac{E}{m}= \frac{r^2-2 M r\pm a\, \sqrt{Mr}}
{r\,\left(r^2-3 M r \pm 2\, a \,\sqrt{M
r}\,\right)^{1/2}}\,\,,
 \label{energy}
\end{equation}
where  $ E= m u_0 $. This expression was first found  and studied in \cite{bardeen}. The vanishing denominator of this expression, determines the radius  $ r_{ph} $  of the limiting photon orbit. That is, the circular motion  exists only in the region  $ r > r_{ph} $. It is not difficult to show that for an extreme Kerr black hole, $ a = M $,  we have $ r_{ph} = M $  for  direct orbits and $ r_{ph} = 4 M $  for retrograde orbits, while for $ a=0 $, we find that $ r_{ph}= 3 M $.

To describe the epicyclic motion,  we introduce a  deviation vector
\begin{equation}
\xi^{\mu}(s)= x^{\mu}(s)  - z^{\mu}(s) \,,
\label{expan}
\end{equation}
and  expand equation (\ref{eqmot}) in powers of  $ \xi^{\mu}(s) $  about the circular orbits. Focusing on the linear approximation,  it is straightforward to show that the epicyclic motion  amounts  to two decoupled oscillations in the radial and vertical directions (the details  can be found in \cite{ag1, ag2}  and  in recent  works \cite{ap, a1}). For the frequency of radial oscillations,  we have
\begin{equation}
\Omega_{r}^2 = \Omega_{0}^2\, \left( 1-\frac{6 M}{r} -\frac{3
a^2}{r^2} \pm \, 8 a \Omega_{s}
\right),
\label{kerrradf}
\end{equation}
whereas the frequency of vertical  oscillations is given by
\begin{equation}
\Omega_{\theta}^2= \Omega_{0}^2\, \left(1
+\frac{3 a^2}{r^2} \mp \, 4 a \Omega_{s} \right).
\label{kerraxif}
\end{equation}
The stability of the circular motion against small oscillations is determined by the  conditions  $\, \Omega_{r}^2 \geq 0 \,$ and $\,\Omega_{\theta}^2 \geq 0.\, $ From the condition for the radial stability, one can infer the radii of the innermost stable circular orbits (ISCOs). As it was first shown in \cite{bardeen}, for a nonrotating black hole, $ a=0 $, we have $ r_{ms}=6 M $, while for the extreme rotating case, $ a= M $, we find  that $ r_{ms}=M $ for the direct orbit and $ r_{ms}= 9 M $ for the retrograde one. Meanwhile, it is easy to verify that the  expression in (\ref{kerraxif}) is always nonnegative in the region of existence and  radial stability of the circular motion. That is, the motion is stable with respect to small oscillations  in the vertical direction.

To summarize, the description of  the circular motion and its small perturbations in the Kerr field  inevitably results in  three  different fundamental frequencies, $ \Omega_0 $,  $ \Omega_r $  and  $ \Omega_{\theta} \,$, unlike the case of Newtonian gravity, where all the three frequencies turn out to be  the same as the  Kepler frequency, $ \Omega_0 = \Omega_r = \Omega_{\theta}=  \Omega_s\,$. (We note that for the  Schwarzschild black hole two of these different frequencies are different. Namely,  $ \Omega_r $ and   $ \Omega_{\theta} = \Omega_0 = \Omega_s  $). It is  the regime of strong gravity near the Kerr black hole, combined with its rotational dynamics, that distinguishes between the three  fundamental frequencies. Therefore, the values of these frequencies at the radii of physical interest are of great importance   for astrophysical implications of the Kerr black hole.

In the following, we  use the coordinate frequency $ \nu $ in physical units, instead of the angular frequency $ \Omega = 2\pi \nu $,  as well as  the characteristic  frequency scale  $ \nu_l=c^3/2\pi G M\simeq 3.2 \cdot 10^4 \left(M_{\odot}/M\right) Hz\,. $ We recall that here $ c $ is the speed of light, $ G $ is the  gravitational constant and  $ M_{\odot} $ is the mass of the Sun. With this in mind and  using expression (\ref{orbf}), we find that the orbital frequency around an extreme  rotating black hole $ (a=M) $ is
\begin{equation}
\nu_0 = \frac{1}{2} \, \nu_l \, \simeq 1.6 \cdot 10^4 \left(\frac{M_{\odot}}{M}\right) Hz\,
\label{dirorbital}
\end{equation}
for  the  direct ISCO whereas, it is  given by
\begin{equation}
\nu_0 = \frac{1}{26} \, \nu_l \, \simeq 1.2 \cdot 10^3 \left(\frac{M_{\odot}}{M}\right) Hz\,
\label{retorbital}
\end{equation}
for the retrograde ISCO. We have performed a detailed  numerical analysis of the orbital and vertical frequencies at  ISCOs around a Kerr black hole with mass $ M= 10 M_{\odot} $. The results of calculations are given in Table I.
\begin{table}[!p]
\renewcommand{\baselinestretch}{1}\normalsize
  \centering
\caption{Orbital and vertical frequencies at ISCOs $ (M= 10 M_{\odot}) $}
\label{}
\begin{tabular}{|c|cccc||cccc|} \hline
&\multicolumn{4}{c||}{direct orbits}&\multicolumn{4}{c|}{retrograde orbits}\\
\hline
\rule[-3mm]{0cm}{9mm}~$a/M$~&$r_{ms}/M$&$\nu_{0}(Hz)$&$\nu_{\theta}(Hz)$&$\nu_{0}/\nu_{\theta}$
&$r_{ms}/M$&$\nu_{0}(Hz)$&$\nu_{\theta}(Hz)$&$\nu_{0}/\nu_{\theta}$\\
\hline
0.00 & 6.00 & 217.73 & 217.73 & 1.00  & 6.00 & 217.73 & 217.73 & 1.00\\
0.10 & 5.67 & 235.32 & 231.91 & 1.01  & 6.32 & 202.54 & 205.15 & 0.99\\
0.20 & 5.33 & 255.93 & 248.03 & 1.03  & 6.64 & 189.28 & 193.91 & 0.98\\
0.30 & 4.98 & 280.49 & 266.52 & 1.05  & 6.95 & 177.59 & 183.79 & 0.97\\
0.40 & 4.61 & 310.32 & 287.97 & 1.08  & 7.25 & 167.20 & 174.65 & 0.96\\
0.50 & 4.23 & 347.48 & 313.16 & 1.11  & 7.55 & 157.92 & 166.33 & 0.95\\
0.60 & 3.83 & 395.41 & 343.20 & 1.15  & 7.85 & 149.56 & 158.74 & 0.94\\
0.70 & 3.39 & 460.42 & 379.58 & 1.21  & 8.14 & 142.02 & 151.81 & 0.94\\
0.80 & 2.91 & 556.00 & 423.99 & 1.31  & 8.43 & 135.12 & 145.38 & 0.93\\
0.90 & 2.32 & 721.41 & 474.68 & 1.52  & 8.72 & 128.84 & 139.47 & 0.92\\
0.99 & 1.24 & 1348.05 & 304.86 & 4.42 & 8.99 & 123.19 & 134.10 & 0.92\\
\hline
\end{tabular}
\end{table}
We note that as the rotation parameter of the black hole grows, the radius of  the direct ISCO  moves towards the event horizon and the associated  orbital  frequency   increases, approaching  its maximum value in (\ref{dirorbital}).
The vertical epicyclic frequency attains its maximum value and then decreases to zero, i.e.  $ \nu_{\theta} =0 $  for $ a=M,\,\, r=M $. The ratio of the frequencies,  $ \nu_{0}/\nu_{\theta}  $,   essentially differs from unity only for the fast enough rotation of the black hole. For the retrograde motion both frequencies decrease with the growth of the rotation parameter (see also Eq. (\ref{retorbital})), whereas their ratio remains  about unity.

It is also of interest to calculate all the three frequencies at direct stable orbits beyond  ISCO ($ r> r_{ISCO} $) around an extreme rotating black hole, $ a=M $. The results of numerical calculations are presented in Table II.
\begin{table}[!p]
\renewcommand{\baselinestretch}{1}\normalsize
 \centering
\caption{Frequencies at direct stable orbits $(r> r_{ISCO}\,,\, a= M= 10 M_{\odot})$}
\label{radfreqt}
\begin{tabular}{|+c|^c^c^c|^c^c^c|}
\hline
\rule[-3mm]{0cm}{9mm}$r/M$ ~&~$\nu_r$(Hz)~&~$\nu_\theta$(Hz)~&~$\nu_0$(Hz)~&~$\nu_\theta/\nu_r$&$\nu_0/\nu_r$&$\nu_0/\nu_\theta~$ \\
\hline
      2.00 &     234.08 &     484.35 &     835.85 &       2.07 &       3.57 &       1.73\\
      2.30 &     244.60 &     462.27 &     712.99 &       1.89 &       2.91 &       1.54\\
      2.90 &     237.48 &     398.45 &     538.85 &       1.68 &       2.27 &       1.35\\
      3.50 &     217.25 &     337.58 &     423.96 &       1.55 &       1.95 &       1.26\\
      3.80 &     206.12 &     311.02 &     380.61 &       1.51 &       1.85 &       1.22\\
      4.10 &     195.14 &     287.13 &     344.02 &       1.47 &       1.76 &       1.20\\
\rowstyle{\bfseries}%
      4.30 &     188.05 &     272.60 &     322.69 &       1.45 &       1.72 &       1.18\\
      4.60 &     177.86 &     252.71 &     294.50 &       1.42 &       1.66 &       1.17\\
      4.90 &     168.28 &     234.89 &     270.12 &       1.40 &       1.61 &       1.15\\
\rowstyle{\bfseries}%
      5.00 &     165.23 &     229.37 &     262.72 &       1.39 &       1.59 &       1.15\\
      5.30 &     156.49 &     213.94 &     242.40 &       1.37 &       1.55 &       1.13\\
      5.60 &     148.36 &     200.05 &     224.53 &       1.35 &       1.51 &       1.12\\
      5.90 &     140.81 &     187.51 &     208.73 &       1.33 &       1.48 &       1.11\\
      6.20 &     133.80 &     176.17 &     194.67 &       1.32 &       1.45 &       1.11\\
      6.50 &     127.30 &     165.87 &     182.11 &       1.30 &       1.43 &       1.10\\
      6.80 &     121.27 &     156.50 &     170.83 &       1.29 &       1.41 &       1.09\\
\rowstyle{\bfseries}%
      7.00 &     117.48 &     150.71 &     163.93 &       1.28 &       1.40 &       1.09\\
\hline
\end{tabular}
\end{table}
We observe that at particular  orbits near the black hole, the predicted values of the epicyclic frequencies (bolded in the table) are close to the corresponding frequencies of twin peaks QPOs, which  are in a 3:2 ratio and have been detected in some black hole binaries: For instance, to {(184, 276 Hz)} for X-ray binary XTE J1550-564; to {(165, 241 Hz)} for X-ray binary H1743-322; to {(113, 168 Hz)} for X-ray binary GRS 1915 + 105.\\

\subsection{The  highest epicyclic frequencies}

It is interesting  that there exist  the highest  epicyclic frequencies of small oscillations  around  circular orbits in the Kerr field. This fact  was first noted in \cite{kato1},  for the frequency of radial oscillations  given by expression (\ref{kerrradf}). Evaluating the first derivative of  this expression  with respect to $ r $, we obtain  the equation
 \begin{equation}
r^3 \left(8 M -r\right) + a^2 \left(5 r^2- 4 M r \right) \pm  2 a \sqrt{M r }\left[a^2 + r \left(M - 6 r\right)\right]= 0\,\,.
\label{maxradfr}
\end{equation}
This equation  determines the radii for both direct and retrograde orbits, at which the radial epicyclic frequency attains its highest value. For the Schwarzschild case, $ a=0 $, it follows that $ r_{max} = 8 M $ and  the associated  frequency  $ \nu_{r(max)} \simeq 707.1 \left(M_{\odot}/{M}\right) Hz \,. $ Meanwhile, for  $ a\neq 0 $, equation (\ref{maxradfr}) can be solved only numerically. In particular, for  $ a =  M $ and for  the direct orbit, we find that $ r_{max} \simeq   2.4 M $ and
\begin{equation}
\nu_{r(max)} \simeq 2453 \left(\frac{M_{\odot}}{M}\right) Hz\,.
\end{equation}
Similarly, for the retrograde orbits at $ a =  M $, we have $ r_{max} \simeq   11.8 M $ and
\begin{equation}
\nu_{r(max)} \simeq 422.6 \left(\frac{M_{\odot}}{M}\right) Hz\,.
\end{equation}

In Figure 1 we plot the dependence of the radial  epicyclic frequency on the radii of circular orbits  around the  Kerr black hole  for different values of the rotation parameter and for $ M= 10 M_{\odot}$.
\begin{figure}[!htmbp]%
\centering%
\includegraphics[width=14cm]{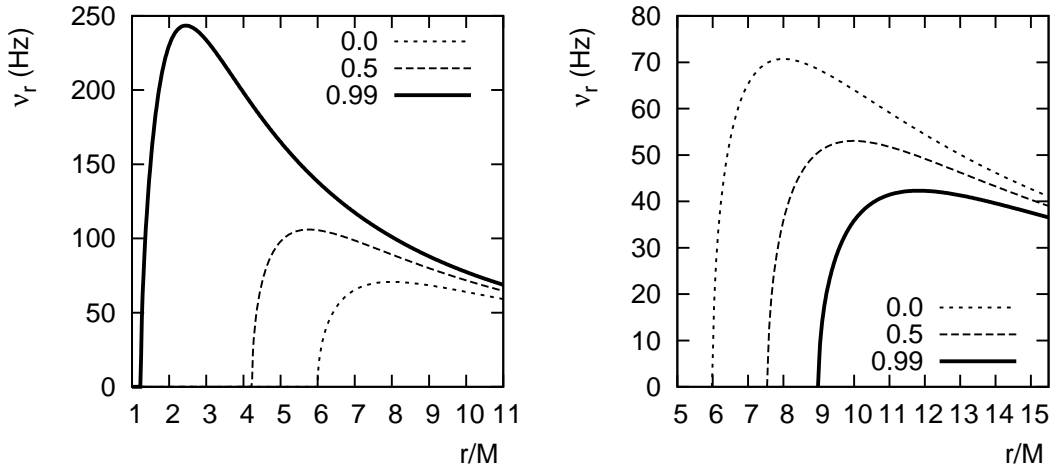}%
\caption{Radial epicyclic frequencies with three values of the rotation parameter $ a/M=0\,,\, 0.5 $ and $ 0.99 \,$. ({\it Left}\,: For direct orbits. {\it Right}\,: For retrograde orbits).}%
\label{freq-rad}%
\end{figure}
The full numerical analysis of equation (\ref{maxradfr}) and the associated values of the radial, vertical and orbital frequencies along with their corresponding ratios are given in Tables III and IV.

Comparing Figure 1 and Table III, we see that for  the direct orbits, with increasing rotation parameter of the black hole, the maxima  of the radial epicyclic frequency shifts towards the event horizon and in the limiting case $ a= M $, the frequency attains its highest value in the near-horizon region.  The accompanying  vertical and orbital frequencies at the same radii also increase to their highest values for $ a= M $. It is also interesting to note that the characteristic  ratios $ \nu_{\theta} : \nu_{r(max)}= 2 : 1  $,~  $ \nu_{0} : \nu_{r(max)}= 2 : 1  $ and  $ \nu_{0} : \nu_{\theta}= 1 : 1  $  remain  almost unchanged  up to large enough values of the rotation parameter. However, for  ${ a \rightarrow M }$ we have the approximate ratios $ \nu_{\theta} : \nu_{r}= 9 : 5  $,~  $ \nu_{0} : \nu_{r}= 5 : 2  $ and  $ \nu_{0} : \nu_{\theta}= 3 : 2 \, $.  {\it Thus, we conclude that at the characteristic stable circular orbits, where  the radial epicyclic frequency attains its highest value,  the ratio $ \nu_{\theta} : \nu_{r}= 2 : 1  $ remains  nearly the  same even for} $ a \rightarrow M $. Remarkably,  this fact is in good enough agreement with the observed twin QPOs frequencies in the X-ray spectrum of some  black hole binaries. For instance, for  $ a \simeq (0.8$--$0.9) M $, the detected pair {(164, 328 Hz)}  in the source GRS 1915 + 105 falls in the expected ranges of the radial $ \nu_r $  and vertical $ \nu_{\theta} $ epicyclic frequencies given in Table~\ref{maxrfreqd}.
Meanwhile, Figure 1 and Table IV for the retrograde orbits show that the highest value of the radial frequency at $ a=0 $ as well as the associated values of the vertical and orbital frequencies decrease with the growth of the rotation parameter and attain their characteristic values  in the limiting case $ a= M $. It is also interesting to note that  the frequencies exhibit, to a good enough accuracy, the ratios $ \nu_{\theta} : \nu_{r(max)}= 2 : 1  $,~  $ \nu_{0} : \nu_{r(max)}= 2 : 1  $ and  $ \nu_{0} : \nu_{\theta}= 1 : 1  $ .

\begin{table}[!p]
\renewcommand{\baselinestretch}{1}\normalsize
  \centering
\caption{The highest radial frequency and the associated vertical and orbital frequencies  at direct orbits $ (M= 10 M_{\odot})$}
\label{maxrfreqd}
\hspace{-0.2 in}\begin{tabular}{|c|ccccccc|} \hline
\rule[-3mm]{0cm}{9mm}~$a/M$~&~$r_{max}/M$~&~$\nu_{r} $(Hz)~&~$\nu_{\theta} $(Hz)~&~$\nu_{0} $(Hz)~&~$\nu_{\theta}/\nu_{r}$~&~$\nu_{0}/\nu_{r}$~&~$\nu_{0}/\nu_{\theta}$~\\
\hline
0.00 & 8.00 & 70.71 & 141.42 & 141.42 & 2.00 & 2.00 & 1.00\\
0.10 & 7.58 & 75.73 & 151.17 & 152.60 & 2.00 & 2.02 & 1.01\\
0.20 & 7.15 & 81.52 & 162.37 & 165.68 & 1.99 & 2.03 & 1.02\\
0.30 & 6.70 & 88.27 & 175.45 & 181.26 & 1.99 & 2.05 & 1.03\\
0.40 & 6.24 & 96.29 & 190.74 & 199.97 & 1.98 & 2.08 & 1.05\\
0.50 & 5.76 & 105.99 & 209.23 & 223.29 & 1.97 & 2.11 & 1.07\\
0.60 & 5.26 & 118.03 & 231.78 & 252.90 & 1.96 & 2.14 & 1.09\\
0.70 & 4.71 & 133.53 & 260.59 & 292.71 & 1.95 & 2.19 & 1.12\\
0.80 & 4.11 & 154.57 & 298.90 & 349.83 & 1.93 & 2.26 & 1.17\\
0.90 & 3.42 & 185.95 & 353.95 & 442.93 & 1.90 & 2.38 & 1.25\\
0.99 & 2.45 & 243.45 & 447.52 & 662.11 & 1.84 & 2.72 & 1.48\\
1.00 & 2.42 & 245.34 & 450.40 & 671.62 & 1.84 & 2.74 & 1.49\\
\hline
\end{tabular}
\end{table}

\begin{table}[!p]
\renewcommand{\baselinestretch}{1}\normalsize
\centering
\caption{The highest radial frequency and the associated vertical and orbital frequencies  at retrograde orbits $(M= 10 M_{\odot})$}
\label{maxrfreqr}
\begin{tabular}{|c|ccccccc|} \hline
\rule[-3mm]{0cm}{9mm}~$a/M$~&~$r_{max}/M$~&~$\nu_{r} $(Hz)~&~$\nu_{\theta} $(Hz)~&~$\nu_{0} $(Hz)~&~$\nu_{\theta}/\nu_{r}$~&~$\nu_{0}/\nu_{r}$~&~$\nu_{0}/\nu_{\theta}$~\\
\hline
0.00 & 8.00 & 70.71& 141.42&141.42 & 2.00& 2.00 & 1.00 \\
0.10 & 8.41 & 66.31& 132.83&131.72 & 2.00& 1.99 & 0.99 \\
0.20 & 8.81 & 62.42& 125.20&123.24 & 2.01& 1.97 & 0.98 \\
0.30 & 9.21 & 58.95& 118.38&115.74 & 2.01& 1.96 & 0.98 \\
0.40 & 9.60 & 55.84& 112.25&109.08 & 2.01& 1.95 & 0.97 \\
0.50 & 9.98 & 53.03& 106.70&103.10 & 2.01& 1.94 & 0.97 \\
0.60 & 10.36& 50.48& 101.65& 97.72 & 2.01& 1.94 & 0.96 \\
0.70 & 10.73& 48.16& 97.04 & 92.85 & 2.02& 1.93 & 0.96 \\
0.80 & 11.10& 46.03& 92.81 & 88.41 & 2.02& 1.92 & 0.95 \\
0.90 & 11.47& 44.07& 88.92 & 84.35 & 2.02& 1.91 & 0.95 \\
1.00 & 11.83& 42.26& 85.32 & 80.63 & 2.02& 1.91 & 0.95 \\
\hline
\end{tabular}
\end{table}

We turn now to the expression of the vertical epicyclic frequency given in (\ref{kerraxif}). For direct orbits and for sufficiently large values of the rotation parameter, this frequency  attains its highest value  as well. The radii of  characteristic  orbits, pertaining to the maxima of the vertical  frequency,  are given by the equation
\begin{equation}
r \left[r^3 + a^2 \left(5 r - 2M\right)\right] + 2 a \sqrt{M r }\left(a^2 - 3 r^2\right)= 0\,\,.
\label{maxvertfr}
\end{equation}
Solving this equation numerically for  $ a=M $, we find that $ r_{(max)}\simeq 1.86 M $ and the highest vertical frequency is
\begin{equation}
\nu_{\theta(max)} \simeq 4875 \left(\frac{M_{\odot}}{M}\right) Hz \,,
\end{equation}
whereas, the radial frequency at this radius has the value
\begin{equation}
\nu_{r} \simeq 2236\left(\frac{M_{\odot}}{M}\right) Hz \,.
\end{equation}
It is easy to see  that  the approximate ratio $ \nu_{\theta(max)}: \nu_{r}= 2 : 1  $  holds in this case as well.

In Figure 2 we plot the vertical epicyclic frequency as a function of the radius of direct orbits, for given values of the rotation parameter and  for $ M= 10 M_{\odot}$  ({\it Left}). Here we also plot the positions of ISCOs and  $ \nu_{\theta(max)} \,$  as the functions of the rotation parameter ({\it Right}).  We see that the  vertical frequency reaches its highest value in the  region of physical interest, $ r_{(max)} > r_{ISCO} $.

\begin{figure}[!htmbp]%
\centering%
\includegraphics[width=14cm]{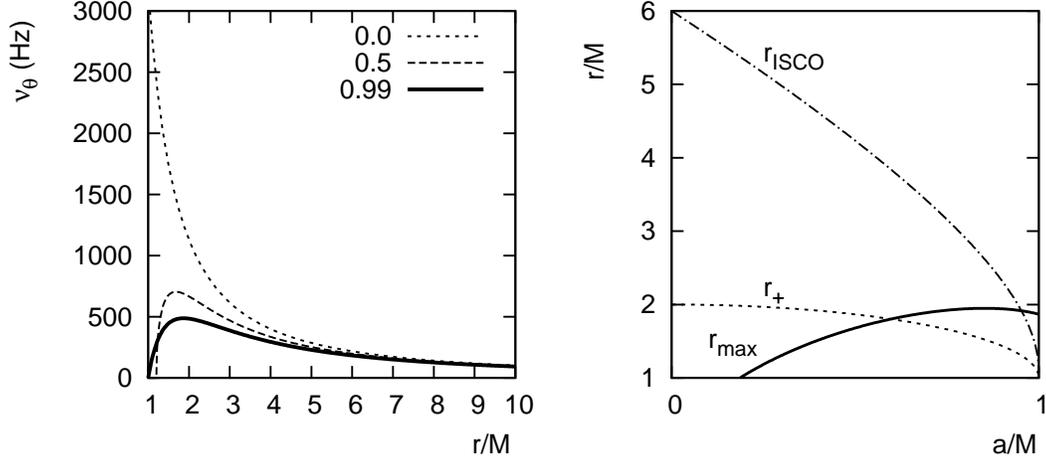}%
\caption{ {Direct orbits}\,: Vertical epicyclic frequencies  with  the rotation parameter $ a/M=0\,, 0.5 $ and $ 0.99 \,$ ({\it Left}). Positions of ISCOs and $ \nu_{\theta(max)} \,$  as the functions of the rotation parameter ({\it Right}).}%
\label{freq-vert}%
\end{figure}

\begin{center}
\section{Geodesic Motion  and Epicyclic Frequencies  around rotating braneworld  black holes}
\end{center}

The measurements of the angular momenta  of  black holes harbored in a number of X-ray binaries reveal high enough values, approaching in some cases
the extreme Kerr limit.  For instance,  the {\it lower bound}  on the rotation parameter of the  black hole in the X-ray binary GRS 1915 + 105  turns out to be  very close to the ({\it upper}) Kerr  bound  for  rotating black holes in general relativity \cite{nara}. Although similar results  are somewhat model-dependent, nevertheless they reinforce the belief  that a black hole  that possesses an angular momentum greater than the Kerr bound could exist. This in turn stimulates the study of the  observable effects of  black holes with over-Kerr bound angular momenta (``over-rotating" black holes). In this section,  we explore this  possibility in an instructive model of geodesic motion around  a rotating  black hole localized on a 3-brane in the Randall-Sundrum scenario \cite{ae2}. Such a black hole  carries a  tidal charge, as an imprint of the extra fifth dimension, and  breaches the Kerr bound for the negative value of the tidal charge. The associated spacetime metric in Boyer-Lindquist coordinates  is given by
\begin{equation}
ds^2  =  -{{\Delta}\over {\Sigma}} \left(dt - a \sin^2\theta\,
 d\phi\,\right)^2 + \Sigma \left(\frac{dr^2}{\Delta} + d\theta^{2}\right) + \frac{\sin^2\theta}{\Sigma}
\left[a dt - (r^2+a^2) \,d\phi \right]^2 \,,
\label{ktmetric}
\end{equation}
where
\begin{eqnarray}
\Delta &=& r^2 + a^2 -2 M r +\beta\,,~~~~~~
\Sigma= r^2+a^2 \cos^2\theta \,.
\label{kmfunc}
\end{eqnarray}
Here $ M $  and $ a $  are the mass and rotation parameters, respectively and  the parameter $ \beta $ is the tidal charge. This metric is the reminiscent of the usual Kerr-Newman metric in which the electric charge of the black hole is superseded by the tidal charge.  Solving  the equation $ \Delta =  0 $, we find that the radius of the event horizon is given by
\begin{equation}
r_{+}= M + \sqrt{M^2 - a^2- \beta}\,,
\label{horizon1}
\end{equation}
which implies the  relation
\begin{equation}
M^2 \geq a^2 +\beta\,,
\label{extreme}
\end{equation}
where the equality holds in the extreme case. We see that  {\it for the  positive tidal charge, the  angular momentum of the braneworld black hole does not exceeds  its mass, whereas  for the negative tidal charge it  is greater than the mass}. The detailed analysis of the properties of metric (\ref{ktmetric}) can be found in \cite{ae2,ap}.

The circular motion of  test particles in metric (\ref{ktmetric}) occurs  in the equatorial plane $ \theta=\pi/2 $, whose  orbital  frequency $\, \Omega_0\,$  has the same form as given in  (\ref{orbf}), where the Kepler frequency is  now given by $ \Omega_s= \left(M r - \beta\right)^{1/2}/r^2\,$. As for the energy of the test particles, it can  be found either by substituting  in equation (\ref{eqmot}) the nonvanishing components of the Christoffel symbols  and repeating  all the steps leading to equation  (\ref{energy}) or simply replacing  the square of the electric charge by the
tidal charge, in the corresponding expression for the usual Kerr-Newman metric (see, for instance \cite{ag1}). As a result, we have
\begin{equation}
\frac{E}{m}= \frac{r^2-2 M r+\beta \pm a\, \sqrt{Mr-\beta}}
{r\,\left[r^2-3 M r+2 \beta \pm 2\, a \,\sqrt{M
r-\beta}\,\right]^{1/2}}\,\,.
\label{orbft}
\end{equation}

The vanishing denominator of this expression  gives the radius of the limiting photon orbit. For $ \beta=M^2 $, the limiting radius is the same as for an extreme Reissner-N\"{o}rdstrom black hole, $ r_{ph}=  M $  $ (r_{+}=M) $. Meanwhile, for the negative tidal charge $\,\beta=-M^2 \,$ $ ({a=\sqrt{2} M}\,,~r_{+}= M)  $, the limiting radii are   $ r_{ph}=  M $ for the the direct motion and $ r_{ph} = 4.82 M  $ for the retrograde one \cite{ae2}.

As for the epicyclic motion, it is frequency in the radial direction is given by
\begin{equation}
\Omega_r^2 =  \frac{\Omega_0^2}{M r - \beta}\left[M r \left(1-\frac{6M}{r}-\frac{3a^2}{r^2}+\frac{9\beta}{r^2}\right)+\frac{4\beta}{r^2}(a^2-\beta) \pm 8 a \Omega_s(Mr-\beta)\right],
\label{bwradfreq}
\end{equation}
while, for  the frequency in the vertical direction we have
\begin{equation}
{\Omega_\theta}^2={\Omega_0}^2\left[1+\frac{a^2}{r^2}\left(1+\frac{2Mr-\beta}{Mr-\beta}\right)\mp 2a\Omega_s\frac{2Mr-\beta}{Mr-\beta}\right].
\label{bwvertfreq}
\end{equation}
We note that these expressions were also given in \cite{ap} and they can be obtained from the general expressions for  the Kerr-Newman field, earlier found in \cite{ag1}, in which  one needs to change the square of the electric charge to the tidal charge, keeping in mind that  the latter may have both positive and negative values. For the vanishing  tidal charge, $\beta =0 $, these expressions go over into those given  in (\ref{kerrradf}) and (\ref{kerraxif}). The boundaries of  the ISCOs in the radial  direction are governed by the equation $ \,\Omega_r^2 =0 \,$. The  analysis of this equation shows that for $ \beta=M^2 $, the limiting radius is ${r_{ms}= 4 M}\,$  just as  for an  extreme Reissner-N\"{o}rdstrom   black hole, whereas for an  over-rotating black hole with $ \beta = -M^2 \,$ $(a=\sqrt{2} M) $  we have  $ r_{ms}= M $  for the direct motion and $ r_{ms}\simeq 11.25M $ for the retrograde one (see \cite{ae2, ap} for details).  Meanwhile, the analysis of expression (\ref{bwvertfreq}) shows that the vertical motion is stable against linear perturbations in this direction.

It turns out that the radial epicyclic frequency in (\ref{bwradfreq}) has maxima at some characteristic orbits,  just as expression (\ref{kerrradf}) for  the Kerr field. In what follows, we will focus on this case. Assuming first that the black hole has a small positive tidal charge, we compute all three fundamental frequencies and their corresponding ratios at direct orbits, for which the radial frequency in (\ref{bwradfreq}) attains its maximum value. The numerical results are summarized in Table~\ref{q01maxrad}.

\begin{table}[!p]
\renewcommand{\baselinestretch}{1}\normalsize
\centering
\caption{The highest radial frequency and the associated vertical and orbital frequencies for the positive tidal charge}
\label{q01maxrad}
\begin{tabular}{|c|ccccccc|} \hline
\multicolumn{8}{|c|}{direct orbits: ~$ \beta=0.1M^2,$~~$ M = 10 M_{\odot}$}\\
\hline
\rule[-3mm]{0cm}{9mm}~$a/M$~&~$r_{max}/M$~&~$\nu_{r} $(Hz)~&~$\nu_{\theta} $(Hz)~&~$\nu_{0} $(Hz)~&~$\nu_{\theta}/\nu_{r}$~&~$\nu_{0}/\nu_{r}$~&~$\nu_{0}/\nu_{\theta}$~\\
\hline
      0.00 &       7.81 &      72.78 &     145.69 &     145.69 &       2.00 &       2.00 &       1.00\\
      0.09 &       7.41 &      77.83 &     155.54 &     156.99 &       2.00 &       2.02 &       1.01\\
      0.19 &       6.99 &      83.65 &     166.83 &     170.17 &       1.99 &       2.03 &       1.02\\
      0.28 &       6.56 &      90.42 &     179.91 &     185.77 &       1.99 &       2.05 &       1.03\\
      0.38 &       6.12 &      98.43 &     195.30 &     204.58 &       1.98 &       2.08 &       1.05\\
      0.47 &       5.66 &     108.10 &     213.71 &     227.80 &       1.98 &       2.11 &       1.07\\
      0.57 &       5.17 &     120.05 &     236.29 &     257.38 &       1.97 &       2.14 &       1.09\\
      0.66 &       4.64 &     135.38 &     264.85 &     296.79 &       1.96 &       2.19 &       1.12\\
      0.76 &       4.07 &     156.08 &     302.77 &     353.12 &       1.94 &       2.26 &       1.17\\
      0.85 &       3.39 &     186.74 &     357.31 &     444.84 &       1.91 &       2.38 &       1.24\\
      0.95 &       2.42 &     244.13 &     452.24 &     666.79 &       1.85 &       2.73 &       1.47\\
\hline
\end{tabular}
\end{table}

Comparing these results with those given in Table~\ref{maxrfreqd}, we see that the observed  pair of frequencies {(164, 328 Hz)}  in the source GRS 1915 + 105 falls in the range of the radial $ \nu_r $  and vertical $ \nu_{\theta} $  frequencies that  corresponds to less values of the rotation parameter, $ a \simeq (0.7$--$0.8) M $. On the other hand, recent observations  give  the lower bound  $  a > 0.98 M $ on the rotation parameter  of the black hole in GRS 1915 + 105 \cite{nara}. Nevertheless, this discrepancy still might not be enough on its own  to judge about the observational appearance (or non-appearance) of the positive tidal charge as the reported constraint on the  bound  of the rotation parameter is highly model-dependent. In some other cases, this constraint is found to be much below (see, for instance, \cite{middle, blum}). The reason for this is that the measurements of the rotation parameter of the black hole are obtained by modeling the thermal X-ray continuum of its accretion disk spectra to infer the inner radius of disk with subsequently identifying it with that of ISCO, which depends only on the mass and rotation parameter of the black hole. However, the accretion disk spectra are generally complex being  accompanied by a radiative tail of non-thermal  nature as well as by the imprints of relativistic effects. Depending on the model under consideration, these  may have a large impact on the estimate of the rotation parameter.

Meanwhile, there exists a fundamental constraint on the values of the positive tidal charge and the rotation parameter, governed by equation (\ref{horizon1}), i.e. by the very existence of the black hole. It follows from this equation that with the growth of the positive tidal charge, the rotation parameter  must decrease and for the large enough  charge,$ \,\beta\rightarrow M^2\,$, the rotation parameter tends to zero. Furthermore, performing  the above  numerical analysis of the epicyclic frequencies  for the tidal charge $\,\beta=0.99 M^2 \,$,  which implies the highest rotation parameter $\, a=0.1 M \,$, we find  the pair of frequencies {($ \nu_r,  \nu_{\theta} $ )}= {(118, 249 Hz). This pair lies far enough away from  the detected pair of frequencies {(164, 328 Hz)} in the source GRS 1915 + 105.
On the basis of all that has been said above, we can conclude that {\it  at least, the large enough positive tidal charge is not supported by the observations of QPOs in black hole systems}.

Next, we suppose that the black hole  possesses the negative tidal charge $\,\beta=-M^2 \,$ and again calculate all three frequencies at the characteristic direct orbits, at which the radial epicyclic frequency attains its maxima.  The results are given in Table~\ref{q-1maxrad}.
\begin{table}[!p]
\renewcommand{\baselinestretch}{1}\normalsize
\centering
\caption{The highest radial frequency and the associated vertical and orbital frequencies for the negative tidal charge}
\label{q-1maxrad}
\begin{tabular}{|c|ccccccc|} \hline
\multicolumn{8}{|c|}{direct orbits: ~$ \beta=-M^2,$~~$ M= 10 M_{\odot}$}\\
\hline
\rule[-3mm]{0cm}{9mm}~$a/M$~&~$r_{max}/M$~&~$\nu_{r} $(Hz)~&~$\nu_{\theta} $(Hz)~&~$\nu_{0} $(Hz)~&~$\nu_{\theta}/\nu_{r}$~&~$\nu_{0}/\nu_{r}$~&~$\nu_{0}/\nu_{\theta}$~\\
\hline
0.00 & 9.64 & 56.20 & 112.29 & 112.29 & 2.00 & 2.00 & 1.00\\
0.14 & 9.09 & 60.69 & 121.06 & 122.28 & 1.99 & 2.01 & 1.01\\
0.28 & 8.53 & 65.94 & 131.25 & 134.11 & 1.99 & 2.03 & 1.02\\
0.42 & 7.95 & 72.15 & 143.25 & 148.35 & 1.99 & 2.06 & 1.04\\
0.57 & 7.36 & 79.63 & 157.61 & 165.82 & 1.98 & 2.08 & 1.05\\
0.71 & 6.74 & 88.84 & 175.16 & 187.83 & 1.97 & 2.11 & 1.07\\
0.85 & 6.09 & 100.51 & 197.15 & 216.53 & 1.96 & 2.15 & 1.10\\
0.99 & 5.40 & 115.87 & 225.74 & 255.80 & 1.95 & 2.21 & 1.13\\
1.13 & 4.64 & 137.33 & 264.91 & 313.73 & 1.93 & 2.28 & 1.18\\
1.27 & 3.77 & 170.50 & 323.56 & 411.81 & 1.90 & 2.42 & 1.27\\
1.41 & 2.53 & 236.74 & 431.78 & 663.00 & 1.82 & 2.80 & 1.54\\
\hline
\end{tabular}
\end{table}
Comparing these results with those given in Table~\ref{maxrfreqd}, it is easy to see that the negative tidal charge  has an enlarging effect on the radii $ r_{max} $, whereas with increasing the rotation parameter of the black hole, the radii again move towards the event horizon, approaching the limiting value for $ a= \sqrt{2} M $.  We also see the appearance  of the approximate  $ 2 : 1 $ ratio between the vertical and radial epicyclic frequencies.
It is important to note that  in the over-rotating case, $ a \simeq 1.27 M $, the values of the radial $ \nu_r $  and vertical $ \nu_{\theta} $  frequencies and their ratio  are in good enough agreement with the detected pair of frequencies {(164, 328 Hz)} in the source GRS 1915 + 105.

Figure~\ref{freq-rad-bwbh}  displays  the positions of the maxima of the radial epicyclic frequencies  for an extreme Kerr black hole and for an over-rotating braneworld black hole with the negative tidal charge $ \beta=-M^2 $. We see that the epicyclic frequencies  in the field of these black holes are observationally  almost indistinguishable. We note that it is the value of the angular momentum (for instance, reported from independent observations)
that would play a crucial role to distinguish between these two black holes and the over-Kerr bound value of  the angular momentum would  be in favor of the potential existence of  the braneworld black hole in the universe.

\begin{figure}[!htmbp]%
\centering%
\includegraphics[width=9.5cm]{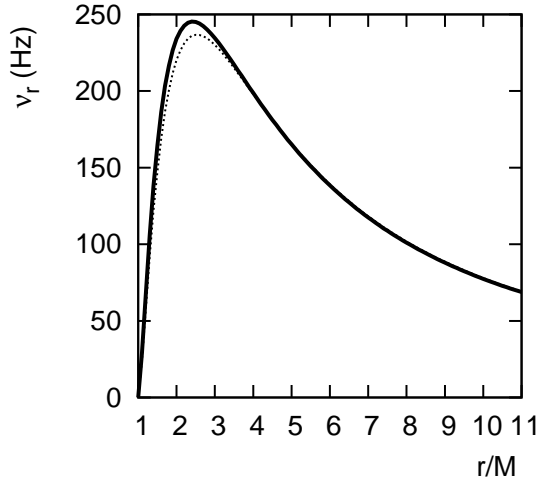}%
\caption{The radial epicyclic frequencies for the direct orbits. The solid line corresponds to the extreme Kerr black hole, $ \beta=0 $,\, $ a= M $ and the dotted line refers to the braneworld black hole, $ \beta=-M^2, $ \,$ a= \sqrt{2} M $.}%
\label{freq-rad-bwbh}%
\end{figure}

\newpage

\section{Conclusion}

In general relativity, the circular motion of test particles around rotating black holes can be described in terms of three fundamental frequencies: The orbital (cyclic)  frequency, the epicyclic frequency of radial oscillations and  the epicyclic frequency of vertical oscillations \cite{ag1,ag2, ag3, ag4}. All three of these  frequencies are different,  encapsulating somewhat  the effects of strong gravity regime near the black holes. Remarkably, the epicyclic frequencies of radial and vertical oscillations can be related to the frequencies of twin peaks high-frequency  QPOs,  discovered  in a number of black hole binaries. The purpose of this paper was  to explore these issues in more detail from the general relativistic point of view. We have considered two instructive models of the geodesic motion; around a Kerr black hole in general relativity and around a rotating  black hole in braneworld gravity. In both cases, we have performed a detailed numerical analysis of the fundamental frequencies at the stable orbits of interest and found the examples  of observationally viable cases of the  epicyclic  frequencies.

As the main result for the Kerr black hole, we have found that for a class of stable orbits, where the radial (or the vertical) epicyclic frequency attains its maxima, the vertical and radial epicyclic frequencies are in an approximate  $ 2 : 1 $ ratio, occurring  even for near-extreme  rotation of the black hole.  We have also found that for the rotation parameter in the range $ a \simeq (0.8$--$0.9) M $, the values of the  epicyclic frequencies underlying this ratio exhibit good enough qualitative agreement with the detected twin peaks frequencies of QPOs in the source GRS 1915 + 105  which, to our best knowledge, appears to be a unique case so far. Furthermore, we have found that at particular class of  orbits near the black hole, the predicted values of the epicyclic frequencies (bolded in Table II) are close to the  frequencies of twin peaks QPOs, which  are in a 3:2 ratio and have been detected in some black hole binaries: For instance, to {(184, 276 Hz)} for X-ray binary XTE J1550-564; to {(165, 241 Hz)} for X-ray binary H1743-322; to {(113, 168 Hz)} for X-ray binary GRS 1915 + 105.  These findings are substantially new. As we noted in the introduction,  the idea of nonlinear resonances that may happen when the epicyclic frequencies  are in a rational relation \cite{ag1} has been utilized in the literature  \cite{abram1, abram2} to develop the models for the observed high frequency QPOs. The results of our paper show that the simple model of pure geodesic motion  can also provide, to some extent,  the observationally viable values for the epicyclic frequencies, without invoking nonlinear resonances between them.

For the rotating braneworld with a tidal charge, we have  performed the full numerical analysis of  the epicyclic frequencies, focusing on the stable orbits, where  the radial epicyclic frequency reaches  its maximum value. Here we have found  that  while the large enough value of the positive tidal charge is not supported by the observations of high frequency QPOs, the situation is different for the negative tidal charge.  For the large enough  value of the negative tidal charge,  the  braneworld black hole exhibits close similarities  with the  Kerr black hole, including the appearance  of the approximate  $ 2 : 1 $ ratio between the vertical and radial epicyclic frequencies. We have also found  that in the case under consideration, the over-rotating braneworld black hole and  the extreme Kerr black hole are  almost indistinguishable in the high-frequency QPOs  observations. We note that these results are also substantially new. On these grounds, we concluded that  the potential existence of the over-rotating  braneworld black hole with the negative tidal charge does not confront with modern observations of the black hole systems in the universe.

\newpage

\section{Acknowledgments}

One of us (A. N. Aliev) thanks  Ekrem  \c{C}alk{\i}l{\i}\c{c} for his  invaluable support and encouragement. The work of G. D. E. is supported by Istanbul University  Scientific Research Project (BAP) No. 3149.


\end{document}